\documentclass[aps,prl,showpacs,preprint,superscriptaddress]{revtex4}

\usepackage{epsfig}

\newcommand{\mpl}{M_{p\overline{\Lambda}}}
\newcommand{\mb}{{M_{\rm bc}}}
\newcommand{\de}{{\Delta{E}}}
\newcommand{\plam}{{p\overline{\Lambda}}}
\newcommand{\plpi}{{p\overline{\Lambda}\pi^-}}
\newcommand{\plg}{{p\overline{\Lambda}\gamma}}

\newcommand{\psigg}{{p\overline{\Sigma}^0\gamma}}

\begin{document}



\epsfysize 25mm \epsfbox{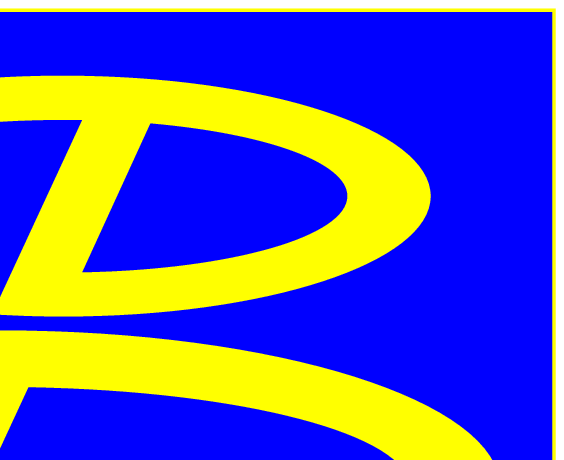}

\begin{flushright}
\vskip -25mm \noindent \hspace*{3.0in}{
                    Belle Preprint 2005-12 \\
                    KEK \ Preprint 2005-3
}
\end{flushright}


\title{ \quad\\[1cm] \Large
Observation of $B^+ \to \plg$}

\tighten

\affiliation{Budker Institute of Nuclear Physics, Novosibirsk}
\affiliation{Chiba University, Chiba}
\affiliation{Chonnam National University, Kwangju}
\affiliation{University of Cincinnati, Cincinnati, Ohio 45221}
\affiliation{Gyeongsang National University, Chinju}
\affiliation{University of Hawaii, Honolulu, Hawaii 96822}
\affiliation{High Energy Accelerator Research Organization (KEK), Tsukuba}
\affiliation{Hiroshima Institute of Technology, Hiroshima}
\affiliation{Institute of High Energy Physics, Chinese Academy of Sciences, Beijing}
\affiliation{Institute of High Energy Physics, Vienna}
\affiliation{Institute for Theoretical and Experimental Physics, Moscow}
\affiliation{J. Stefan Institute, Ljubljana}
\affiliation{Korea University, Seoul}
\affiliation{Kyungpook National University, Taegu}
\affiliation{Swiss Federal Institute of Technology of Lausanne, EPFL, Lausanne}
\affiliation{University of Ljubljana, Ljubljana}
\affiliation{University of Maribor, Maribor}
\affiliation{University of Melbourne, Victoria}
\affiliation{Nagoya University, Nagoya}
\affiliation{Nara Women's University, Nara}
\affiliation{National Central University, Chung-li}
\affiliation{National United University, Miao Li}
\affiliation{Department of Physics, National Taiwan University, Taipei}
\affiliation{H. Niewodniczanski Institute of Nuclear Physics, Krakow}
\affiliation{Nihon Dental College, Niigata}
\affiliation{Niigata University, Niigata}
\affiliation{Osaka City University, Osaka}
\affiliation{Osaka University, Osaka}
\affiliation{Panjab University, Chandigarh}
\affiliation{Peking University, Beijing}
\affiliation{Princeton University, Princeton, New Jersey 08544}
\affiliation{University of Science and Technology of China, Hefei}
\affiliation{Seoul National University, Seoul}
\affiliation{Sungkyunkwan University, Suwon}
\affiliation{University of Sydney, Sydney NSW}
\affiliation{Tata Institute of Fundamental Research, Bombay}
\affiliation{Toho University, Funabashi}
\affiliation{Tohoku Gakuin University, Tagajo}
\affiliation{Tohoku University, Sendai}
\affiliation{Department of Physics, University of Tokyo, Tokyo}
\affiliation{Tokyo Institute of Technology, Tokyo}
\affiliation{Tokyo Metropolitan University, Tokyo}
\affiliation{Tokyo University of Agriculture and Technology, Tokyo}
\affiliation{University of Tsukuba, Tsukuba}
\affiliation{Virginia Polytechnic Institute and State University, Blacksburg, Virginia 24061}
\affiliation{Yonsei University, Seoul}
   \author{Y.-J.~Lee}\affiliation{Department of Physics, National Taiwan University, Taipei} 
    \author{M.-Z.~Wang}\affiliation{Department of Physics, National Taiwan University, Taipei} 
  \author{K.~Abe}\affiliation{High Energy Accelerator Research Organization (KEK), Tsukuba} 
   \author{K.~Abe}\affiliation{Tohoku Gakuin University, Tagajo} 
   \author{H.~Aihara}\affiliation{Department of Physics, University of Tokyo, Tokyo} 
   \author{Y.~Asano}\affiliation{University of Tsukuba, Tsukuba} 
   \author{V.~Aulchenko}\affiliation{Budker Institute of Nuclear Physics, Novosibirsk} 
 \author{T.~Aushev}\affiliation{Institute for Theoretical and Experimental Physics, Moscow} 
   \author{S.~Bahinipati}\affiliation{University of Cincinnati, Cincinnati, Ohio 45221} 
   \author{A.~M.~Bakich}\affiliation{University of Sydney, Sydney NSW} 
% ...
\collaboration{The Belle Collaboration}

\begin{abstract}

We report the first observation of the radiative hyperonic $B$ decay
$B^+ \to \plg$, using a 140 fb$^{-1}$ data sample recorded on the
$\Upsilon({\rm 4S})$ resonance with the Belle detector at the KEKB
asymmetric energy $e^+e^-$ collider. The measured branching
fraction is ${\mathcal B}(B^+ \to \plg) = (2.16 ^{+0.58}_{-0.53}
\pm 0.20) \times 10^{-6}$. We examine its
$\mpl$ distribution and observe a peak near threshold. This feature
is expected by the short-distance
$ b \to s \gamma$ transition.
A search for $B^+ \to \psigg$ yields no
significant signal and we set a 90\% confidence-level upper
limit on the branching fraction 
of ${\mathcal B}(B^+ \to \psigg) < 4.6 \times 10^{-6}$.

\vskip1pc

\end{abstract}
%
%
%

\maketitle
{\renewcommand{\thefootnote}{\fnsymbol{footnote}}

\setcounter{footnote}{0}

\normalsize

%
%


The $ b \to s \gamma$ penguin diagram is responsible for
the large rates of the observed radiative $B \to K^*
\gamma$~\cite{kstarg} decays. It is also a good probe of new
physics beyond the Standard Model~\cite{btosg}. Recently, the Belle
collaboration reported a very stringent limit of ${\mathcal O} (10^{-6})$ on
the branching fraction of two-body $B^+ \to \plam$
decays~\cite{2body} but found an unexpectedly large rate for the
three-body decay $B^0 \to \plpi$~\cite{plpi}, which proceeds,
presumably, via the $b \to s$ penguin process. One interesting
feature of the $B^0 \to \plpi$ decay is that the observed
proton-$\overline{\Lambda}$ mass $M_{\plam}$ spectrum peaks
near threshold. Naively, a suppression of  ${\mathcal O}(\alpha_{EM})$ is
expected for the $B^+ \to \plg$  decay relative to $B^+ \to \plam$
if the former process is bremsstrahlung-like.
In contrast, a short-distance
$ b \to s \gamma$ contribution can lead naturally
to a non-bremsstrahlung-like energetic photon spectrum and an
enhancement of  $M_{\plam}$  at low mass; the former
distribution can be compared to the recently measured $b \to
s\gamma$ inclusive photon energy spectrum~\cite{gammaspec}. These
features motivate our study of $B^+\to\plg$.
Some theoretical
predictions~\cite{theory} for the branching fraction of $B^+ \to
\plg$ are at the $10^{-6}$ level, which is in the
sensitivity range of the B-factories.

We use a data sample of $152\times 10^6$ $B\overline{B}$ pairs,
corresponding to an integrated luminosity of 140
fb$^{-1}$, collected by 
the Belle detector 
at the KEKB~\cite{KEKB} asymmetric energy $e^+e^-$ 
collider. The Belle detector is a large-solid-angle
magnetic spectrometer that consists of a three-layer silicon
vertex detector (SVD), a 50-layer central drift chamber (CDC), an
array of aerogel threshold \v{C}erenkov counters (ACC), a
barrel-like arrangement of time-of-flight scintillation counters
(TOF), and an electromagnetic calorimeter (ECL) comprised of
CsI(Tl) crystals located inside a super-conducting solenoid coil
that provides a 1.5~T magnetic field.  An iron flux-return located
outside of the coil is instrumented to detect $K_L^0$ mesons and
to identify muons.  The detector is described in detail
elsewhere~\cite{Belle}.

To identify the charged tracks, the proton ($L_p$), pion ($L_\pi$)
and kaon ($L_K$) likelihoods are determined from the information
obtained by the hadron
identification system (CDC, ACC and TOF).
Prompt proton candidates must satisfy the requirements of $L_p/(L_p+L_K)> 0.6 $
and $L_p/(L_p+L_{\pi})> 0.6$, and not be associated with the decay
of a $\Lambda$ baryon. The proton selection efficiency is 
about 84\% (88\% for $p$ and 80\% for $\overline{p}$) for particles with
momenta at 2 GeV/$c$, and the fake rate is about
10\% for kaons and 3\% for pions.

The prompt proton candidates are also required
to satisfy track quality criteria based on track impact
parameters relative to the interaction point (IP). The deviations
from the IP position are required to be within 0.3 cm in the
transverse ($x$-$y$) plane, and within $\pm$3 cm in the $z$
direction, where the $z$ axis is opposite the direction of the positron beam.
%
Candidate ${\Lambda}$ baryons are reconstructed from two oppositely
charged tracks, one treated as a proton and the other as a pion,
and must have a mass within $5\sigma$ of the
nominal $\Lambda$ mass, as well as
a displaced vertex and flight direction consistent with a
$\Lambda$ originating from the interaction point. To reduce
background, a $L_p/(L_p+L_{\pi})> 0.6$ requirement is applied to
the proton-like track.
%
%
Photon candidates are selected from  the neutral clusters within
the barrel ECL (with polar angle between $33^\circ$ and
$128^\circ$) having energy greater than 500 MeV.
We discard any photon candidate if the  mass, in combination with any
              other photon above 30 (200) MeV, is within $\pm 18$ ($\pm
              32$) MeV/$c^2$ of the nominal mass of the $\pi^0$ ($\eta$)
              meson.  The above selection criteria are optimized using
              Monte Carlo (MC) simulated event samples.

Candidate $B$ mesons are formed
by combining 
a proton with a $\overline{\Lambda}$ and a photon~\cite{conjugate},
 each defined using the above criteria, and requiring
the beam-energy constrained mass, $\mb =
\sqrt{E^2_{\rm beam}-p^2_B}$, and the energy difference, $\de =
E_B - E_{\rm beam}$,
to lie in the ranges 5.2 GeV/$c^2 < \mb < 5.29$
GeV/$c^2$ and $-0.2$ GeV $ < \de < 0.5$ GeV.
              Here, $p_B$ and $E_B$ refer to the momentum and energy,
              respectively, of the reconstructed $B$ meson, and
              $E_{\rm beam}$ refers to the beam energy, all in the
              $\Upsilon$(4S) rest frame.
Because of the $\de > -0.2$ GeV 
requirement, background from $B$ feed-down is negligible except that 
from $B^+ \to \psigg$ decay 
where $\overline{\Sigma}$ subsequently decays to 
$\overline{\Lambda}\gamma$ almost 100\% of the time. 
The $\psigg$ events can form a nearby
peak (shifted about -100 MeV in $\de$) with respect to the signal peak in 
the $\mb-\de$ region.  

The dominant background for $B^+ \to \plg$ decay is from 
continuum $e^+e^- \to q\bar{q}$
processes, where $q = u, d, s, c$. 
The continuum background is  
evaluated with an MC sample of 120 million
continuum events. In the $\Upsilon({\rm 4S})$ rest frame,
continuum events are jet-like while $B\overline{B}$ events are
spherical. We follow the scheme defined in Ref.~\cite{etapk} and
combine seven shape variables to form a Fisher
discriminant~\cite{fisher} in order to maximize the distinction
between continuum processes and signal. The variables used have
almost no correlation with $\mb$ and $\de$. Probability density
functions (PDFs) for the Fisher discriminant and the cosine of the
angle between the $B$ flight direction and the beam direction in
the $\Upsilon({\rm 4S})$ frame are combined to form the signal
(background) likelihood ${\cal L}_s$ (${\cal L}_b$). We require the
likelihood ratio ${\cal R} = {\cal L}_s/({\cal L}_
s+{\cal L}_b)$ to be greater than 0.75; this suppresses
about 86\% of the background while retaining 78\% of the signal.
The optimal selection requirement is determined by maximizing
$N_s/\sqrt{N_s+N_b}$, where $N_s$ and $N_b$ denote the expected number of
signal and background events; here a signal branching fraction of
$4\times10^{-6}$ is assumed.


\begin{figure}[t!]
\epsfig{file=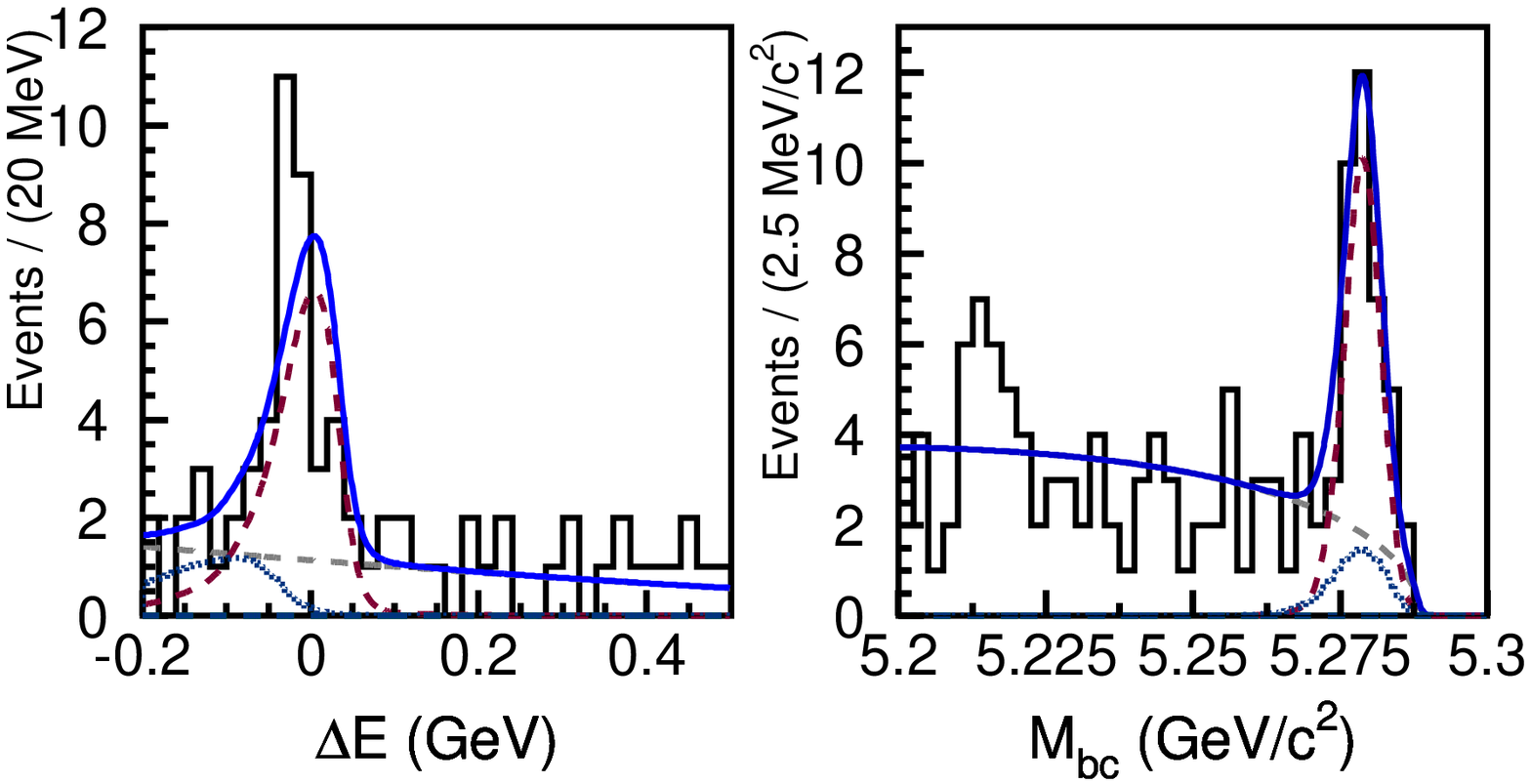,width=3.3in} \caption{The distributions of
$\de$ (for $M_{\rm bc} > 5.27\,{\rm GeV}/c^2$) and $\mb$ (for
$-0.135\,{\rm GeV} < \Delta E < 0.074\,{\rm GeV}$) for $B^0 \to
\plg$ candidates having $M_{p\overline{\Lambda}} < 2.4\,{\rm GeV}/c^2$.
The solid, light dashed and dark dashed lines represent the combined fit
result, fitted background and fitted signal, respectively. 
The dotted lines represent projections
of 10 assumed $\psigg$ events for comparison. }
\label{fg:plgmbde}
\end{figure}

We perform an unbinned extended maximum likelihood fit to the
events with $-0.2$ GeV$<\de<$ 0.5 GeV and $\mb>$ 5.2 GeV/$c^2$ in order to
determine the signal yield, $\Sigma$ feed-down, and $q\bar{q}$ background. 
The extended likelihood function is
defined as
\[\small
 {\cal L} =\frac{e^{-(N_\Lambda+N_\Sigma+N_{q\bar{q}})}}{N!}\prod_{i=1}^{N}
\left[\mathstrut^{\mathstrut}_{\mathstrut}
 N_\Lambda P_\Lambda(M_{{\rm bc}_i},\Delta{E}_i)\right.\]
\[\left.\mathstrut^{\mathstrut}_{\mathstrut}+N_\Sigma
P_\Sigma(M_{{\rm bc}_i},\Delta{E}_i)+
N_{q\bar{q}}P_{q\bar{q}}(M_{{\rm bc}_i},\Delta{E}_i)\right],
\]
where $N$ is the total number of events in the 
fit; $P_\Lambda$, $P_\Sigma$, and $P_{q\bar{q}}$ are the
PDFs for $\plg$, $\psigg$, and continuum background, respectively; 
$N_\Lambda$, $N_\Sigma$, and $N_{q\bar{q}}$ are the corresponding number
of candidates.

The $\plg$ and $\psigg$ PDFs are
two-dimensional functions approximated by smooth histograms  
from MC simulation.
We use the parametrization first suggested by the ARGUS
collaboration~\cite{Argus}, $ f(\mb)\propto \mb\sqrt{1-(\mb/E_{\rm
beam})^2} \exp[-\xi (1-(\mb/E_{\rm beam})^2)]$, to model the
background $\mb$ distribution, and a quadratic polynomial for the
background $\de$ shape. We perform a two-dimensional unbinned fit
to the $\de$ {\it vs} $\mb$ distribution, with the signal and
background normalizations as well as the continuum background
shape parameters allowed to float.


The $\de$ distribution (with $\mb >$ 5.27 GeV/$c^2$) and the $\mb$
distribution (with $-$0.135 GeV$<\de<$ 0.074 GeV) for the region
$\mpl<$ 2.4 GeV/$c^2$ are shown in Fig.~\ref{fg:plgmbde} along with the
projections of the fit.  
The two-dimensional unbinned fit gives a
$B^+\to\plg$ signal yield of $34.1^{+7.1}_{-6.6}$ with a
statistical significance of $8.6$ standard deviations and a
$B^+\to\psigg$ yield of $0.0\pm4.7$. The significance is
defined as $\sqrt{-2{\rm ln}(L_0/L_{\rm max})}$, where $L_0$ and
$L_{\rm max}$ are the likelihood values returned by the fit with
signal yield fixed at zero and its best fit value,
respectively.

We measure the differential branching fraction of $\plg$ by fitting 
the yield in bins of $\mpl$, as shown in Fig.~\ref{fg:phase}, and
correcting for the corresponding detection efficiency as determined
from a large MC sample of events distributed uniformly in phase space.
The results of the fits along with the efficiencies and the partial
branching fractions are given in Table~\ref{bins}. 
In these fits, the signal yields are constrained to be non-negative.
The yield is consistent with null signal for higher $\mpl$ bins if
the non-negative constraint is removed.
The observed mass distribution in Fig.~\ref{fg:phase} peaks at low
$p\overline{\Lambda}$ mass, a feature seen also in $B^0\to\plpi$ and
$B^+\to p\overline{p}K^+$ decays~\cite{plpi,pph}.


\begin{figure}[htb]
\centering \mbox{\epsfig{figure=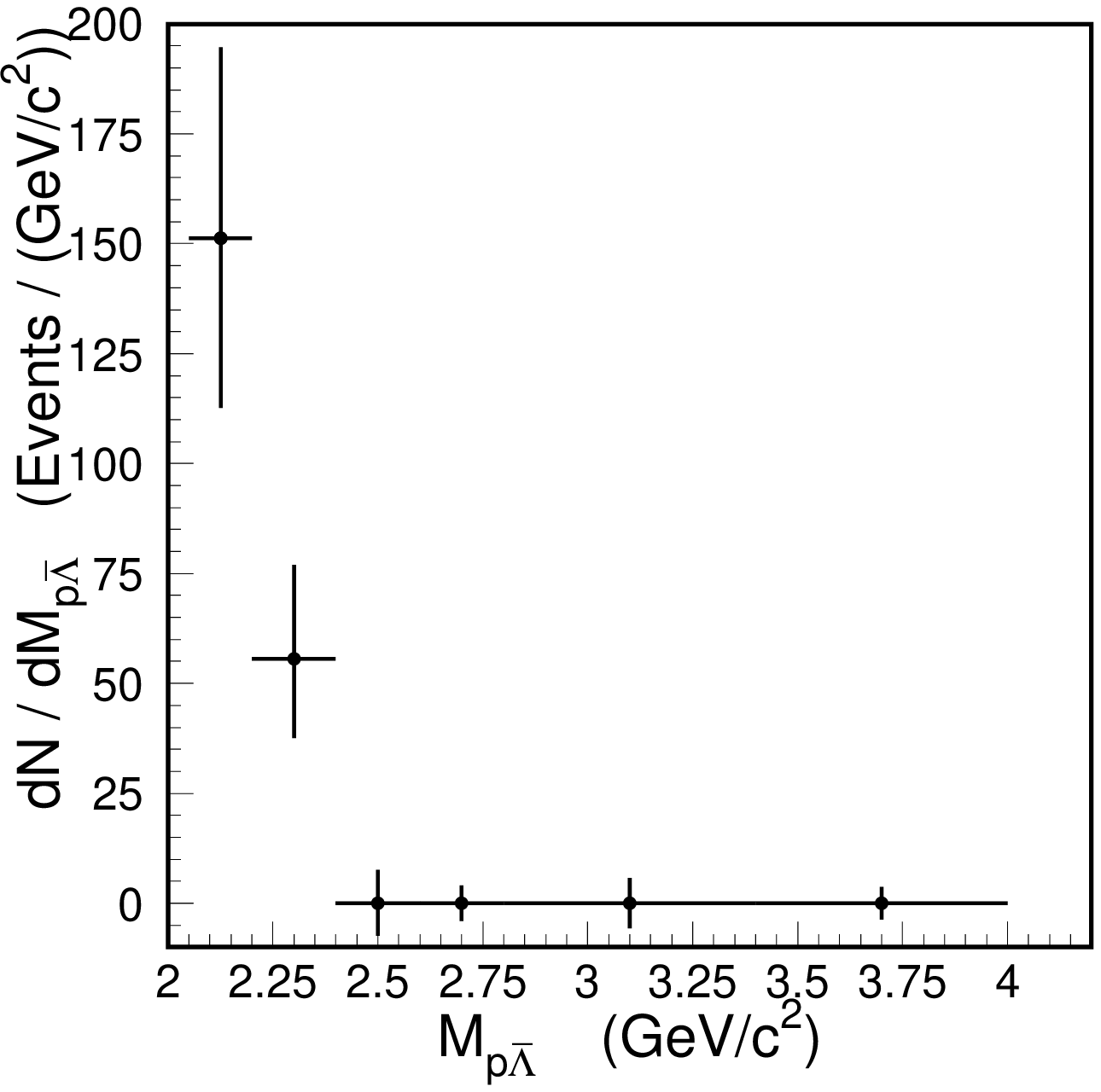,width=3.4in}}
\centering \caption{The differential yield for
$B^0\to \plg$ as a function of $M_{p \overline{\Lambda}}$. 
} 
\label{fg:phase}
\end{figure}

\begin{table}[htb]
\caption{The event yield, efficiency, and branching
fraction (${\cal B}$)
for each  $M_{p\overline{\Lambda}}$ bin.}
\label{bins}
\begin{center}
\begin{tabular}{cccc}
$M_{p\overline{\Lambda}}$ (GeV/$c^2$)& ~~Signal Yield~~&
~~Efficiency(\%)~~& ${\cal B}$ ($10^{-6}$)
\\
\hline $<2.2$& $22.7^{+6.5}_{-5.8}$& 10.6& $1.41^{+0.40}_{-0.36}$
\\
\hline $2.2-2.4$& $11.1^{+4.3}_{-3.6}$& 9.8&
$0.74^{+0.29}_{-0.24}$
\\
\hline $2.4-2.6$& $0.0^{+1.5}_{-1.5}$& 9.3& $0.00^{+0.11}_{-0.11}$
\\
\hline $2.6-2.8$& $0.0^{+0.8}_{-0.8}$& 9.9& $0.00^{+0.06}_{-0.06}$
\\
\hline $2.8-3.4$& $0.0^{+3.4}_{-3.4}$& 9.6& $0.00^{+0.23}_{-0.23}$
\\
\hline $3.4-4.0$& $0.0^{+2.2}_{-2.2}$& 9.6& $0.00^{+0.15}_{-0.15}$
\\
\hline Total & $33.8^{+9.0}_{-8.1}$& - & $2.16^{+0.58}_{-0.53}$
\\
\end{tabular}
\end{center}
\end{table}


We also study the angular distribution of the proton in the baryon pair
system. The angle $\theta_X$ is
measured between the proton direction and the $\gamma$
direction in the baryon pair rest frame. Figure~\ref{fg:thetap}
shows the efficiency corrected $B$ yield in bins of $\cos
\theta_X$. This distribution supports the $b \to s \gamma$
fragmentation picture where the $\Lambda$ tends to emerge
opposite the direction of the photon. 
We define the angular asymmetry as $A = {
{N_{\cos\theta_{X+}} - N_{\cos\theta_{X-}}}\over
{N_{\cos\theta_{X+}} + N_{\cos\theta_{X-}}}}$, where $N_{\cos\theta_{X+}}$ and
$N_{\cos\theta_{X-}}$
stand for the efficiency corrected $B$ yield with $\cos\theta_X > 0$ and
 $\cos\theta_X < 0$, respectively. The measured value for $A$ is 
$0.36^{+0.23}_{-0.20}$.

\begin{figure}[htb!]
\epsfig{file=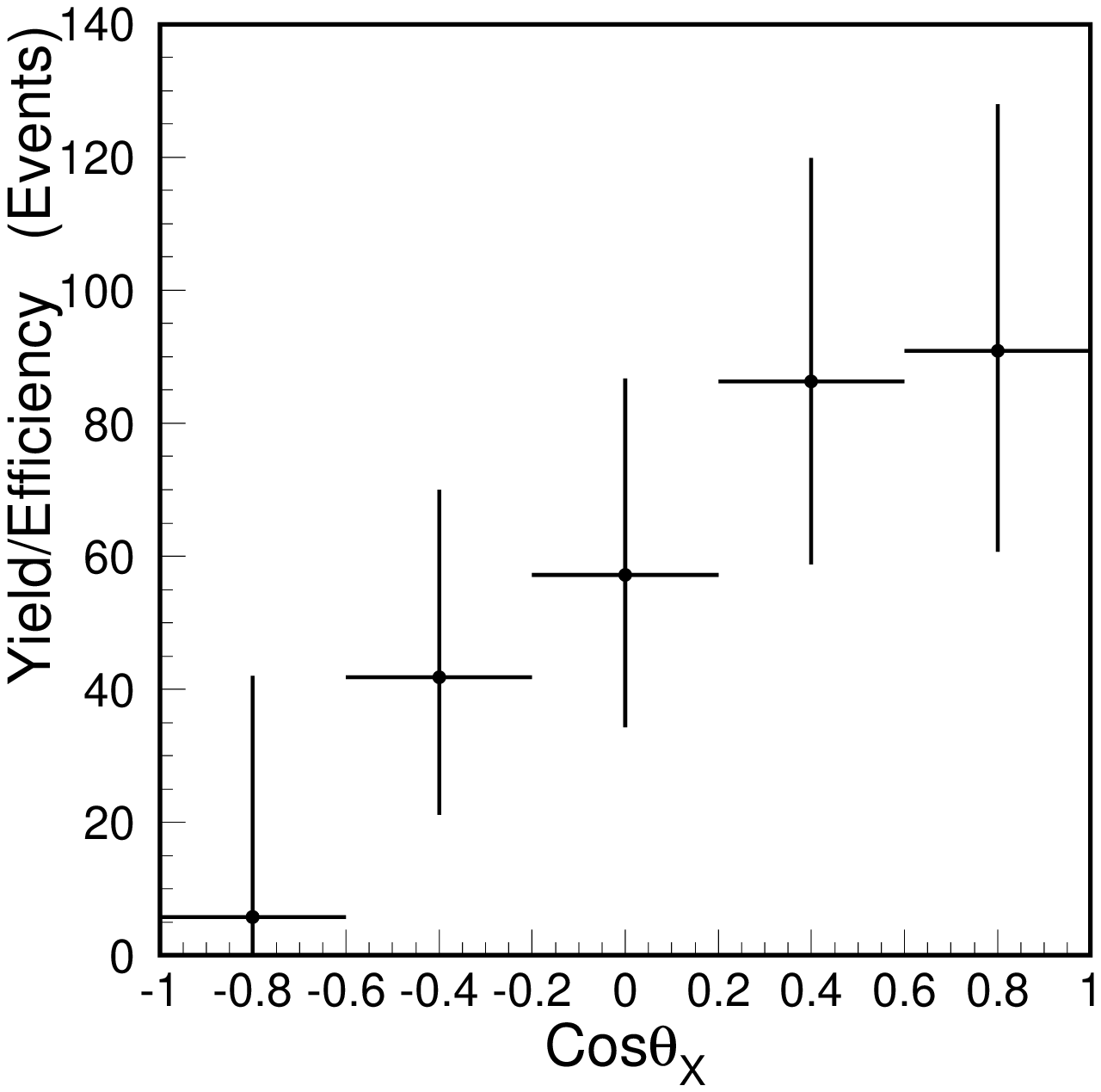,width=3.3in} \caption{Efficiency corrected yield
versus $\cos\theta_X$ in the baryon
pair system. 
} \label{fg:thetap}
\end{figure}

The systematic uncertainty in particle selection is studied using
high statistics control samples. Proton identification is studied
with a  $\Lambda \to p \pi^-$ sample. The tracking efficiency is
studied with a $D^*$ sample, using both full and partial
reconstruction. Based on these studies, we sum the correlated errors
linearly and assign
a 4.1\% error for proton identification
and 4.9\% for the tracking efficiency.

For $\Lambda$ reconstruction, we have an additional uncertainty of 2.5\% on the
efficiency for off-IP track reconstruction, determined from the
difference of $\Lambda$ proper time distributions for data and MC
simulation. 
There is also a 1.2\%
error associated with the $\Lambda$ mass selection and a $0.5\%$ error for the
$\Lambda$ vertex selection~\cite{2body}. Summing the
errors for $\Lambda$ reconstruction, we obtain
a systematic error of 2.8\%.

The 2.2\% uncertainty for the photon detection is determined from radiative
Bhabha events.  
For the $\pi^0$ and
$\eta$ vetoes, we compare the fit results with and without the vetoes;
the difference in the branching fraction is 0.5\%, which is taken
as the associated systematic error.

Continuum suppression is studied by varying the selection
criteria on $\cal R$ in the interval 0 -- 0.9 to see if there is
any systematic trend in the signal fit yield. We
quote a 2.5\% error for this.

The systematic uncertainty from fitting is 2.2\%, which is
determined by assuming uncorrelated $\mb$ and $\de$ PDFs,
and by varying the parameters of
the signal and background PDFs by $\pm 1\sigma$. The MC
statistical uncertainty and modeling with six $\mpl$ bins
contributes a 4.4\% error (obtained by changing the $\mpl$ bin
size). The error on the number
of total $B\overline{B}$ pairs is 0.5\%. 
The error from the sub-decay branching fraction of $\Lambda \to
p\pi^-$ is 0.8\%~\cite{PDG}.

We combine the above uncorrelated errors in quadrature. The total systematic
error is 9.2\%.

We see no evidence for the decay $B^+ \to p\overline{\Sigma}^0\gamma$.
We use the fit results to estimate the expected background, and
compare this  with the observed number of events in the $\psigg$
signal region ($-$0.20 GeV$<\de<$ 0.04 GeV and $\mb >$ 5.27
GeV/$c^2$) in order to set an upper limit on the
yield~\cite{Highland,Gary,Conrad}. 
The estimated
background for $\mpl<4.0$ GeV/$c^2$ is $84.0
\pm 9.2$, the number of observed events is 96, and the
systematic uncertainty is 9.2\%; from these, the upper limit yield is
35.5 at 90\% confidence level.
Assuming the $B^0 \to \psigg$ three-body decay is uniform in phase space, 
the overall efficiency including
the loss from the $\mpl<4.0$ GeV/$c^2$
requirement is 5.1\%;
the 90\% confidence-level upper limit
for the branching fraction is ${\mathcal B}(B^0 \to
\psigg)  < 4.6 \times 10^{-6}$.

In summary, we have performed a search for the radiative baryonic
decays $B^+ \to \plg$, and $\psigg$ with $152$ million $B\overline{B}$
events. A clear signal is seen in the $\plg$ mode, and we measure
a branching fraction of ${\cal B}(B^+ \to \plg)=
{(2.16\,^{+\,0.58}_{-\,0.53} \; ({\rm stat})\pm 0.20 \;
({\rm syst})) \times 10^{-6}}$, which is consistent with the upper
limit set by CLEO\cite{Edwards:2003js}. The yield of the $B^0\to\psigg$ mode
is not statistically significant, and we set the 90\% confidence
level upper limit of
${\mathcal B}(B^0 \to\psigg)  < 4.6 \times 10^{-6}$.

We thank the KEKB group for the excellent operation of the
accelerator, the KEK cryogenics group for the efficient
operation of the solenoid, and the KEK computer group and
the National Institute of Informatics for valuable computing
and Super-SINET network support. We acknowledge support from
the Ministry of Education, Culture, Sports, Science, and
Technology of Japan and the Japan Society for the Promotion
of Science; the Australian Research Council and the
Australian Department of Education, Science and Training;
the National Science Foundation of China under contract
No.~10175071; the Department of Science and Technology of
India; the BK21 program of the Ministry of Education of
Korea and the CHEP SRC program of the Korea Science and
Engineering Foundation; the Polish State Committee for
Scientific Research under contract No.~2P03B 01324; the
Ministry of Science and Technology of the Russian
Federation; the Ministry of Higher Education, Science and 
Technology of the Republic of Slovenia;  
the Swiss National Science Foundation; the National Science 
Council and the Ministry of Education of Taiwan; and 
the U.S. Department of Energy.



\begin{thebibliography}{99}

\bibitem{kstarg}  R. Ammar {\it et al.} (CLEO Collaboration),
              Phys. Rev. Lett. {\bf 71}, 674 (1993);
              B. Aubert {\it et al.} (BaBar Collaboration),
              Phys. Rev. Lett. {\bf 88}, 101805 (2002);
              M. Nakao {\it et al.}(Belle Collaboration),
 Phys. Rev. D {\bf 69},
               112001 (2004).

\bibitem{btosg} T. Hurth, E. Lunghi and W. Porod, Nucl. Phys. B {\bf 704}, 
56 (2005) and references therein.



\bibitem{2body}
K.~Abe {\it et al.} (Belle Collaboration),
Phys. Rev. D {\bf 65}, 091103 (2002).

\bibitem{plpi}
M.-Z.~Wang, {\it et al.}  (Belle Collaboration),
Phys.\ Rev.\ Lett.\  {\bf 90}, 201802 (2003).


\bibitem{gammaspec} P.Koppenburg, {\it et al.} (Belle Collaboration),
 Phys. Rev. Lett. {\bf 93}, 061803 (2004).


\bibitem{theory}
H.Y.~Cheng and K.C.~Yang, Phys. Lett. B {\bf 533}, 271 (2002);
C.Q.~Geng and Y.K.~Hsiao, Phys. Lett. B {\bf 610}, 67 (2005).

\bibitem{KEKB}
S.~Kurokawa and E.~Kikutani,
Nucl. Instr. and Meth. A {\bf 499}, 1 (2003).

\bibitem{Belle}
A.~Abashian {\it et al.} (Belle Collaboration),
Nucl. Instr. and Meth. A {\bf 479}, 117 (2002).


\bibitem{conjugate} {Throughout this report,
inclusion of charge conjugate modes is always implied
unless otherwise stated.}


\bibitem{etapk}{
 K.~Abe {\it et al.} (Belle Collaboration), Phys. Lett. B 
{\bf 517}, 309 (2001).
}


\bibitem{fisher}{
    R.A.~Fisher, Annals of Eugenics {\bf 7}, 179 (1936).
}




\bibitem{Argus}{
    H. Albrecht {\it et al.}, Phys. Lett. B {\bf 241}, 278 (1990);
        {\it ibid}. B {\bf 254}, 288 (1991).
}



\bibitem{pph}{
M.-Z.~Wang {\it et al.} (Belle Collaboration),
Phys. Rev. Lett. {\bf 92}, 131801 (2004).
}



\bibitem{PDG}
S.~Eidelman {\it et al.} (Particle Data Group), Phys. Lett. B {\bf
592}, 1 (2004).


\bibitem{Highland}
R.D. Cousins and V.L. Highland,
Nucl. Instr. and Meth. A {\bf 320}, 331 (1993).


\bibitem{Gary} G.J. Feldman and R.D. Cousins, Phys. Rev. D {\bf 57},
 3873 (1998).

\bibitem{Conrad} J. Conrad {\it et al.},
Phys. Rev. D {\bf 67}, 012002 (2003).

\bibitem{Edwards:2003js}
K.~W.~Edwards {\it et al.}  (CLEO Collaboration),
Phys.\ Rev.\ D {\bf 68}, 011102 (2003) .




\end{thebibliography}
\end{document}